\documentclass[a4paper,11pt]{article}
\usepackage{jinstpub} 
\usepackage{subcaption}

\title{\boldmath ML for the hKLM at the 2nd Detector}

\author[a,b,1]{R. Kelleher \note{Corresponding author.}}
\author[a]{A Vossen}
\affiliation[a]{Duke University,\\Durham, North Carolina, 27708, USA}
\affiliation[b]{University of Michigan,\\Ann Arbor, Michigan, 48109, USA}

\emailAdd{rowankel@umich.edu}

\abstract{The present research applies Graph Neural-Networks (GNNs) for energy measurement and particle identification tasks for a proposed second detector at the future Electron Ion Collider (EIC). In particular, an steel-scintillator sampling calorimeter would provide neutral hadron ($K_L$ and neutron) energy measurements and identification, as well as separation of muons from hadrons. Using detector simulations, particle hits in the detector are represented as graphs, and a GNN is trained for either classification or prediction. Furthermore, we developed a parameterization of the scintillator optical photon simulation that yields a 20-fold speed up compared to the default simulation. We find that the GNN method outperforms classical methods at the same tasks, and we report projections for the energy and timing resolution, and identification accuracy of the calorimeter. We also present an integration of the GNN method into a Multi-Objective Optimization framework, enabled by an automated pipeline of data generation, GNN training, and detector performance evaluation. We utilize the optimization to quantify the tradeoffs between different performance metrics at high and low energies when changing the detector design parameters, such as the steel/scintillator thickness.}


\begin{document}
\maketitle
\flushbottom

\section{Introduction}
This research studies a potential hadronic calorimeter for the second detector at the EIC~\cite{kelleher2026designexpectedperformancehklm}. This detector would consist of alternating steel and scintillator layers, similar to the proposed CORE detector~\cite{CORE} and the Belle II KLM detector~\cite{Wang:2003bm}. In addition to $K_L$ and muon identification (MuID), the detector would function as a hadronic calorimeter for neutral hadrons, primarily $K_L$ and neutrons.
A key feature of this research is the utilization of machine learning techniques for each part of the study: Normalizing flows (NF) are used to improve the scintillator optical photon simulation; GNNs use the low-level detector response to perform particle identification and calorimetry; and bayesian optimization is used to investigate the tradeoffs between competing performance metrics. 

\begin{figure}
    \centering
    \includegraphics[width=0.55\linewidth]{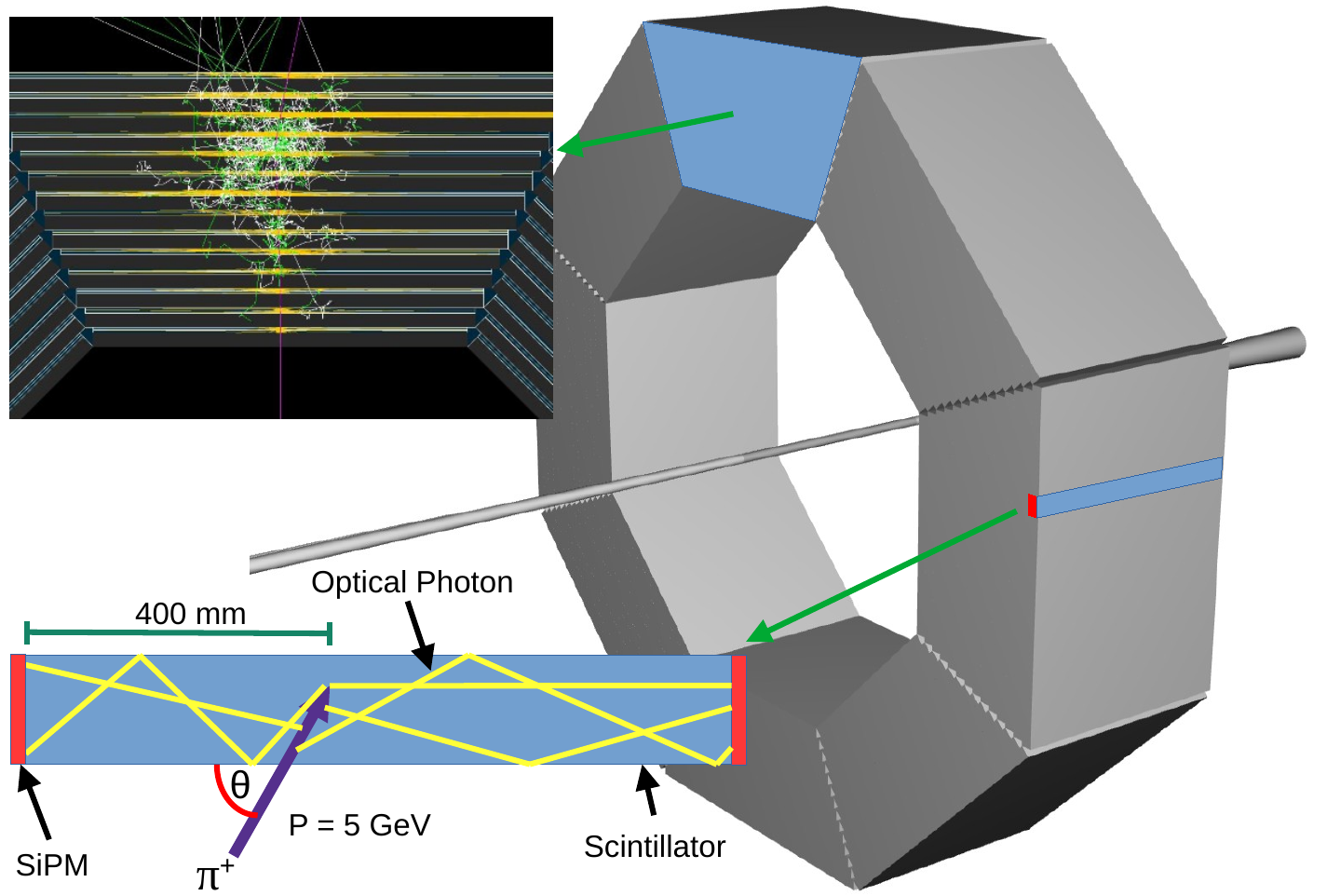}
    \includegraphics[width=0.42\linewidth]{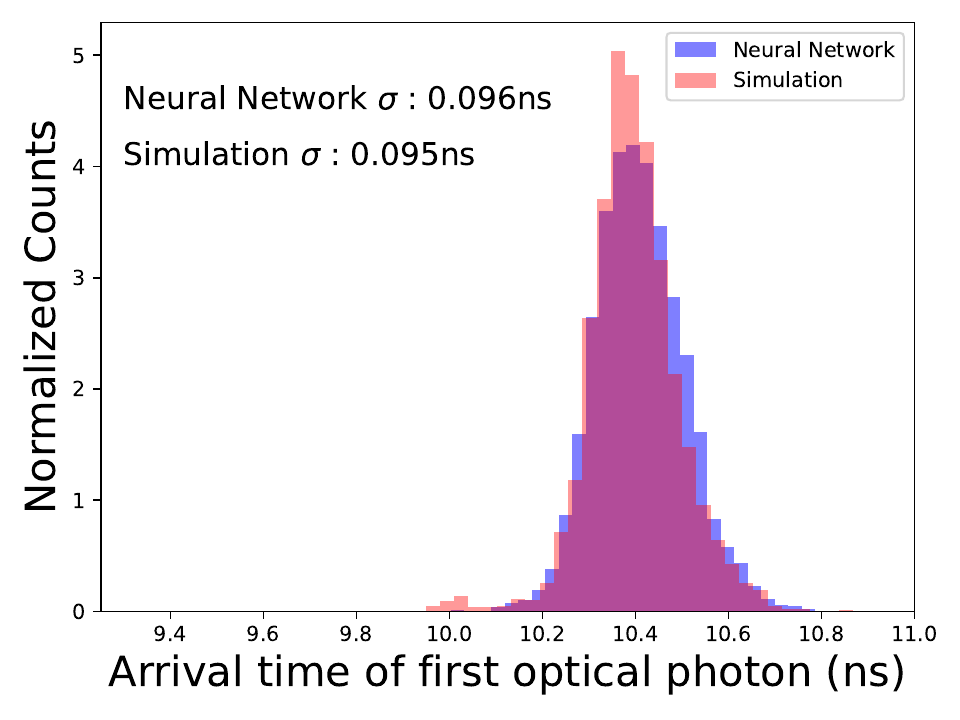}
    \caption{Left: rendering of hKLM around the EIC beam pipe. An example charged pion event is shown in a single sector, and an individual scintillator is highlighted, with a diagram showing the inputs to the NF inference. Right: first optical photon arrival times for 5,000 events with the GEANT4 simulation (red) and sampled from the NF model (blue).}
    \label{fig:rendering}
\end{figure}

The hKLM is a barrel detector with 8 staves forming an octagon around the beam pipe, as shown in the left panel of Figure~\ref{fig:rendering}. The nominal design for the detector has 5.55 cm of steel and 2.00 cm of scintillator per layer, for 14 total layers. Section~\ref{sec:optimization} describes the ML-aided optimization of the material thicknesses and layer count. The scintillator layers are segmented into bars that run parallel to the beam pipe with Silicon photo-multipliers (SiPM) capping the bars for light collection and signal readout.

\section{Simulations}
We utilize the Detector Design for High Energy Physics (DD4HEP)~\cite{frank_markus_2018_1464634} framework for detector simulations. DD4HEP builds the detector geometry from a compact file and runs GEANT4 with a particle gun to simulate the detector response. Although GEANT4 provides an optical photon simulation, we implement a faster optical photon parameterization, producing a speed up of approximately 20-fold. Our method estimates the photon yield based on the hit position and energy deposition, and uses a NF model~\cite{normalizingflows} to sample photon arrival times. We train a NF model on approximately 1 million events from the full GEANT4 optical photon simulation to transform photon arrival times from the truth distribution to a normal distribution. For inference, the trained model is reversed: samples are drawn from a normal distribution and transformed via the inverse of the model.
The model is conditioned on the charged particle position, angle, and momentum. We find that the NF model succeeds at modeling timing distribution produced by the simulation for full timing distribution, as well as the first photon timing distribution, show in the right panel of Figure~\ref{fig:rendering}.

\section{Reconstruction and Performance}
\begin{figure}
    \centering
    \includegraphics[width=0.51\linewidth]{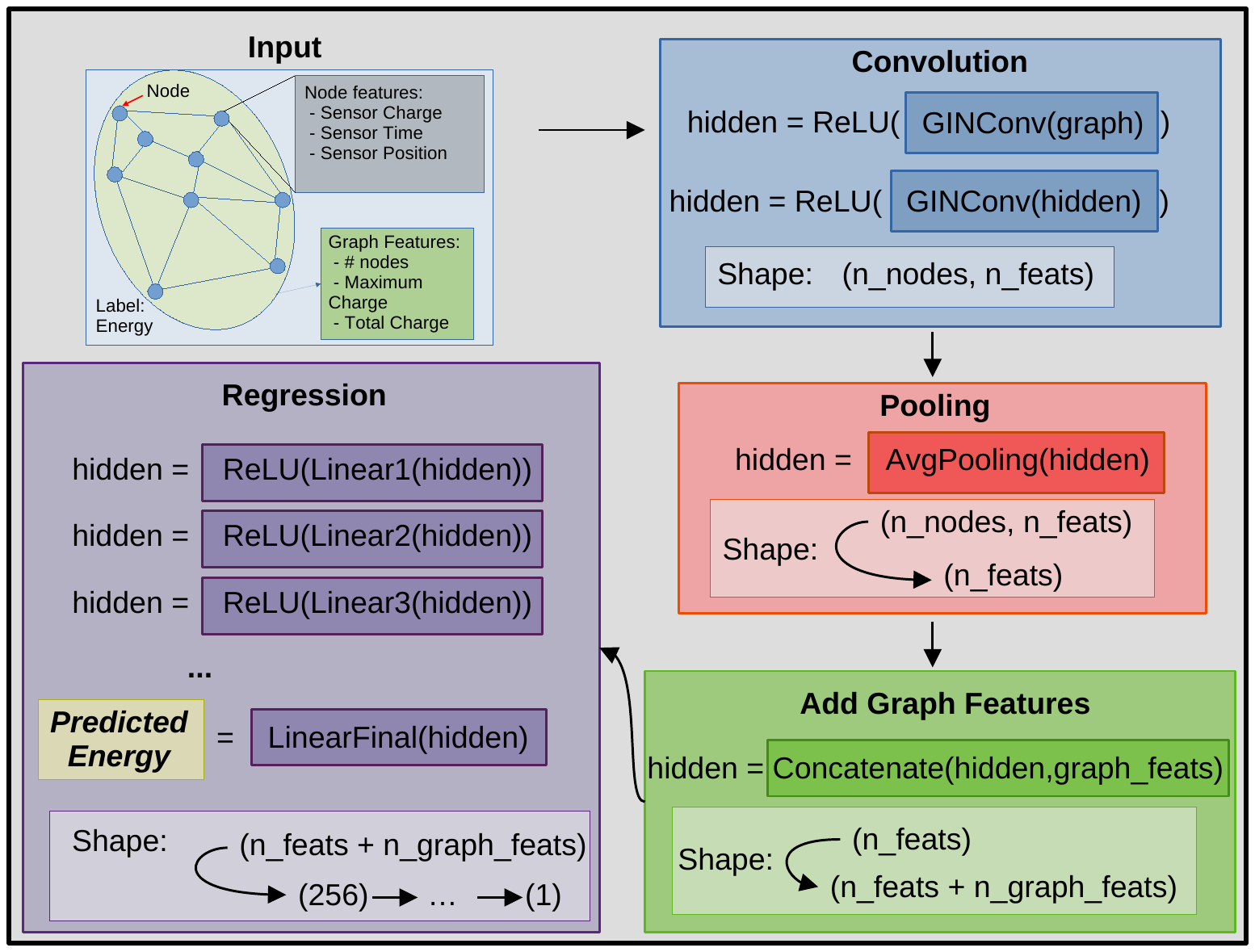}
    \includegraphics[width=0.45\linewidth]{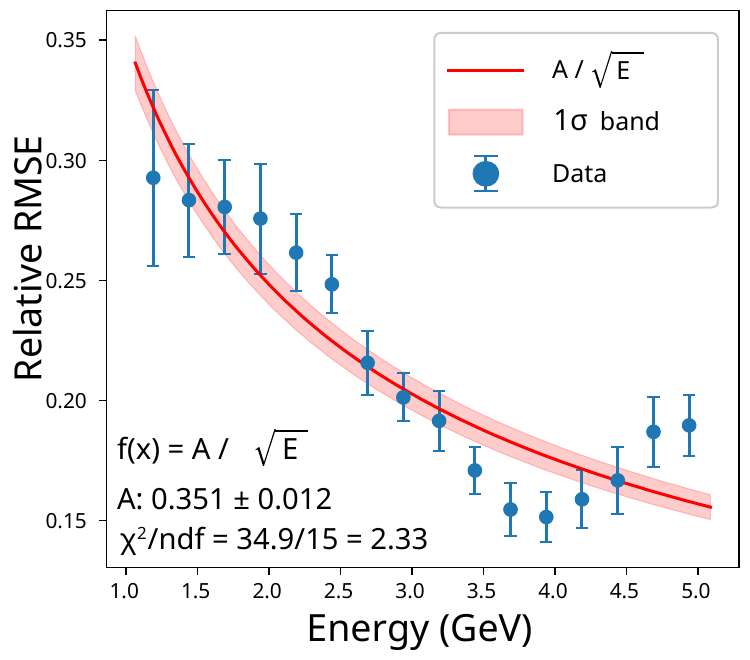}
    \caption{Left: GNN architecture, with input graph structure in top left. Right: GNN energy prediction relative error $vs.$ particle energy (blue points). RMSE is defined as $\mathrm{RMSE} \equiv \sqrt{(\langle E_\mathrm{pred} - E_\mathrm{true})^2\rangle}$ for each true energy bin. The relative RMSE is the ratio of the RMSE to the true energy bin center. Error bars are estimated by training 30 models and taking standard deviation. The fit (red line) produces a resolution of 35.1$\pm$1.2\%/$\sqrt{E}$.}
    \label{fig:GNN_arch}
\end{figure}

The structure of the detector naturally lends itself to a graph structure, where each SiPM, e.g the sensor highlighted in the left panel of Fig.~\ref{fig:rendering}, is represented as a node. Each node is described by features (the hit time and charge), and its position within the detector in cartesian coordinates $(x,y,z)$. The shower shape in the detector is encoded in the graph by including only the SiPMs that exceed a three photon threshold. The nodes are connected using a k-nearest-neighbors algorithm with $k = 6$; the largest graphs have more than 100 nodes, whereas the smallest has 2. The GNN architecture is illustrated in the left panel of  Figure~\ref{fig:GNN_arch}.
The final layer is adjusted for the task.

A sample of 25,000 particles, shot from a particle gun at varying polar and azimuthal angles, is used for reconstruction. For neutral hadron energy measurements, neutrons are used. For MuID, independent samples of pions and muons are generated, each with 25,000 examples. The particles are generated with momenta between 0.5~GeV/$c$ and 5.0~GeV/$c$, polar angle in the range $70^\circ < \theta < 110^\circ$, and azimuthal angle in the range $0^\circ \le \phi \le 360^\circ$. 70\% of the sample is used for training the GNN, with 15\% reserved for validation throughout the training process. The GNN is trained for a maximum 100 epochs, and training is stopped early if validation loss increased three epochs in a row. The final 15\% of the sample is used for generating the results shown in Figure~\ref{fig:GNN_arch} (right). We estimate a resolution of (35.1 $\pm$ 1.2)\%/$\sqrt{E}$, and the fit has $\chi^2/\text{ndf} = 2.33$, which represents a significant improvement over calorimeters with similar alternating steel and scintillator designs. A comparison of the GNN method to a conventional method is detailed in Ref~\cite{kelleher2026designexpectedperformancehklm}.

\begin{figure}
\vspace{-0.3cm}
    \centering
    \includegraphics[width=0.24\linewidth]{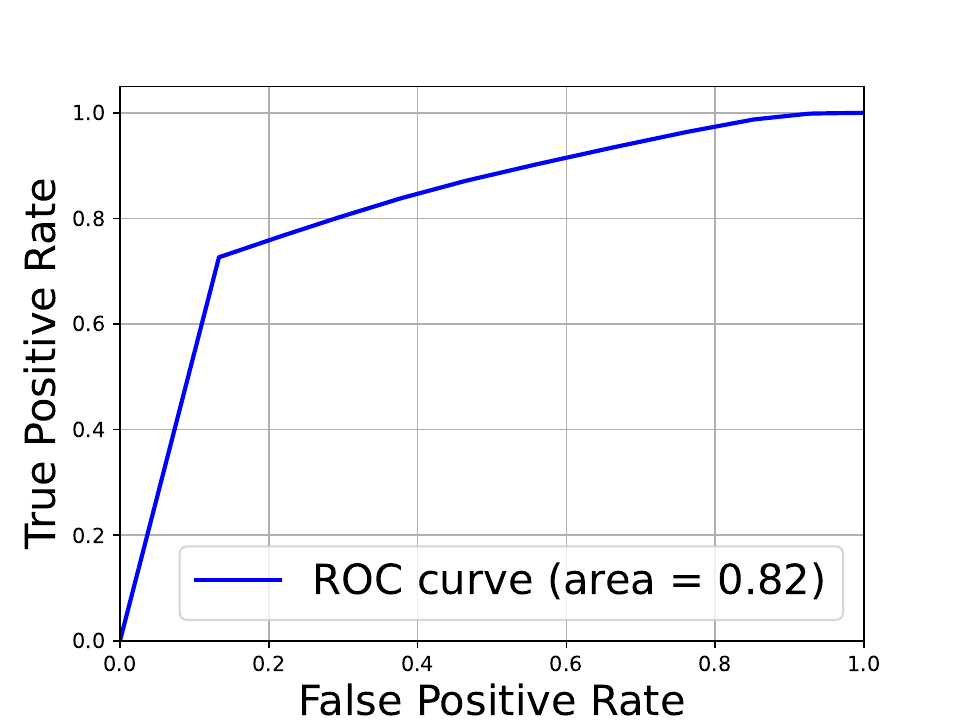}
    \includegraphics[width=0.24\linewidth]{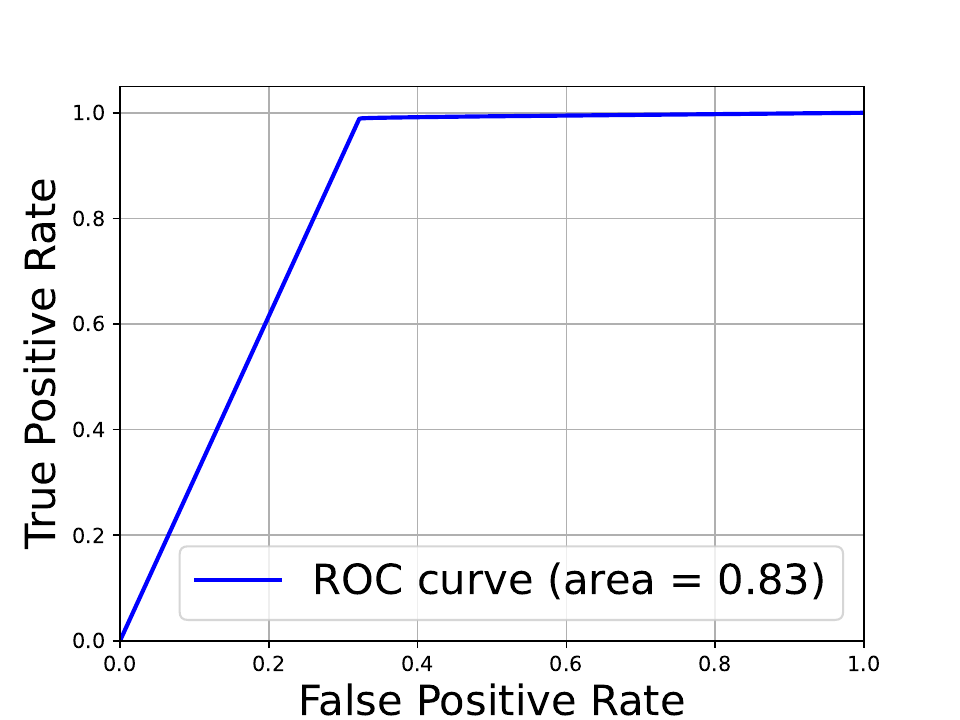}
    \includegraphics[width=0.24\linewidth]{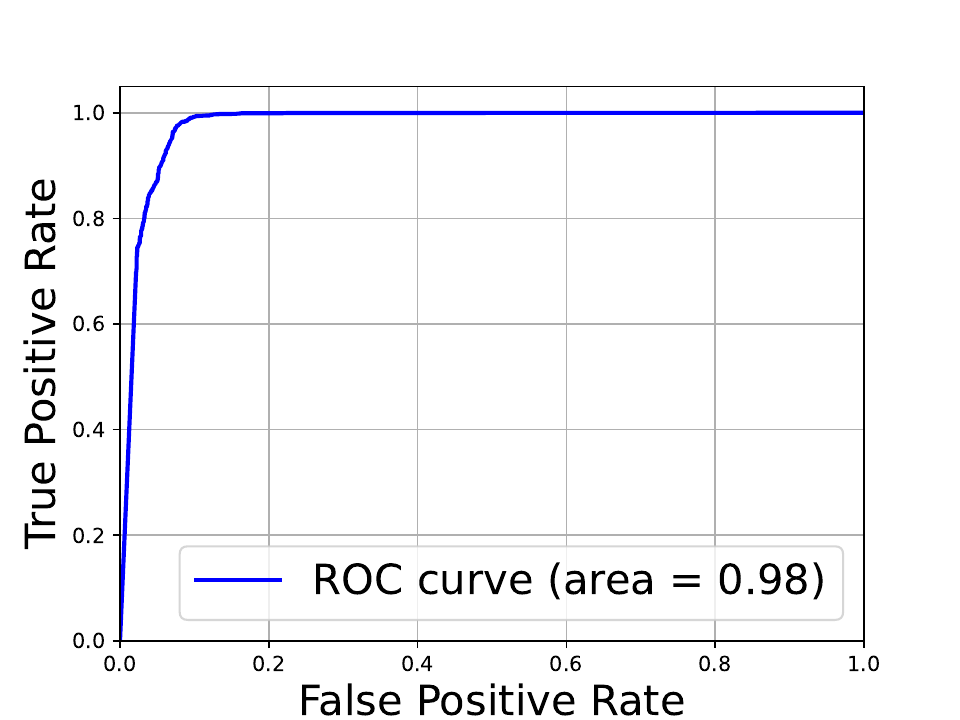}
    \includegraphics[width=0.24\linewidth]{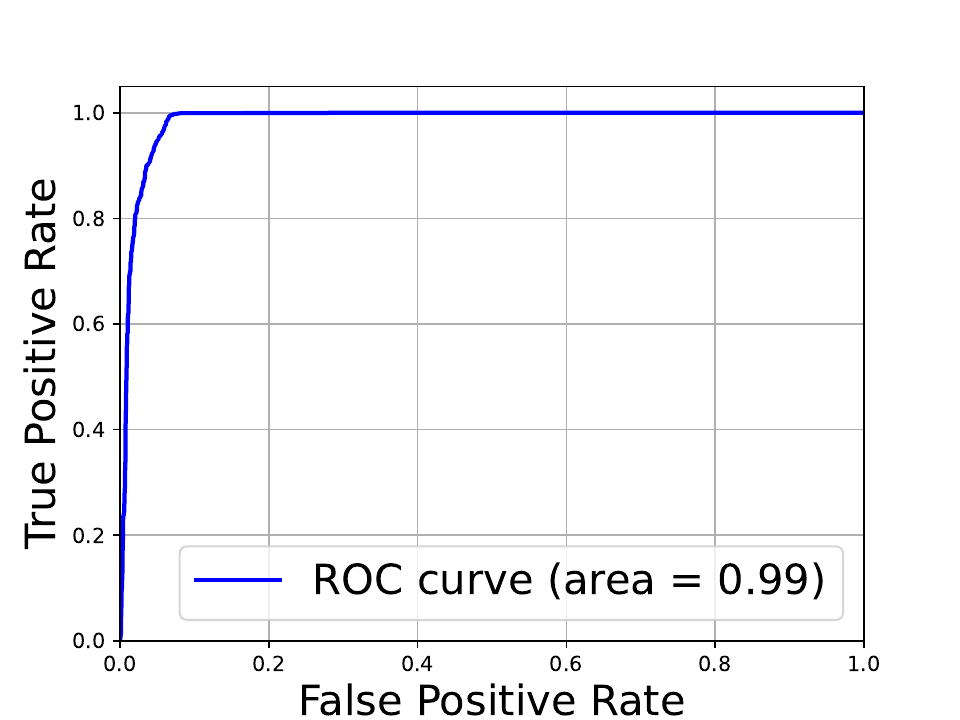}
\caption{ROC curves from left to right: conventional MuID at low and high energies, followed by GNN MuID at low and high energies.}
    \label{fig:results}
\end{figure}

We apply the same training strategy for MuID with a combined sample of muons and pions. The area under the receiver operating characteristic (ROC) curve is used as the performance metric. In the right panel of Figure~\ref{fig:results}, we compare the results from a conventional MuID algorithm (left and next to left) to that from the GNN (next to right and right), using two separate momentum bins: a low energy bin containing particles between 0.5~GeV/$c$ and 2.75~GeV/$c$ (left); and a high energy bin containing particles between 2.75~GeV/$c$ and 5~GeV/$c$ (right). The conventional MuID algorithm takes the deepest scintillator layer reached by a particle and assigns the particle a score equal to the muon purity of tracks that stop in that layer $P(L) = \frac{N_\mu(L)}{N_\mu(L) + N_\pi(L)}$, calibrated by a training sample.
\section{Optimization}
\label{sec:optimization}
The MuID and energy prediction performance depend on the hKLM design parameters, in particular the steel thickness, scintillator thickness, and number of layers.
We model the influence of the design parameters on the detector performance with the AID2E framework~\cite{diefenthaler2024ai}, which trains a surrogate model to estimate the performance of a particular design. For each trial, a design is chosen, training data is generated, and two separate GNNs are automatically trained for MuID and energy reconstruction. 

\begin{figure}
    \centering
    \includegraphics[width=0.8\linewidth]{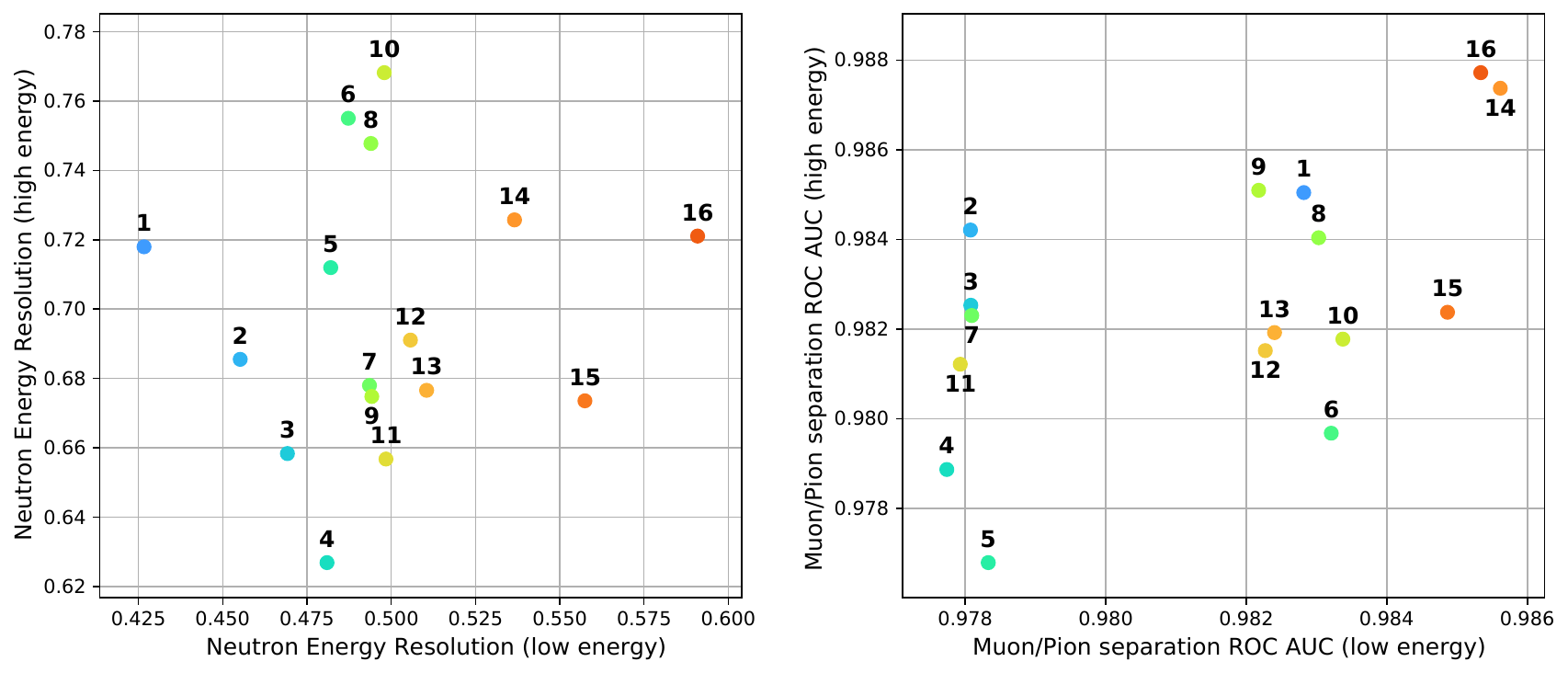}
    \caption{Each panel shows a 2D projection of the 4D Pareto front. Left: neutron predicted energy relative error high $vs.$ low energy performance. Right: MuID high $vs.$ low energy ROC area under the curve.}
    \label{fig:pareto}
\captionof{table}{Geometry parameter values for the trials shown in Figure~\ref{fig:pareto}.}
\label{tab:trials_basic}
\resizebox{\textwidth}{!}{%
\begin{tabular}{l|cccccccccccccccc}
\hline
\textbf{Parameter} & \textbf{1} & \textbf{2} & \textbf{3} & \textbf{4} & \textbf{5} & \textbf{6} & \textbf{7} & \textbf{8} & \textbf{9} & \textbf{10} & \textbf{11} & \textbf{12} & \textbf{13} & \textbf{14} & \textbf{15} & \textbf{16} \\
\hline
\texttt{Number of layers}  & 17 & 17 & 16 & 18 & 18 & 17 & 17 & 16 & 18 & 14 & 18 & 15 & 18 & 16 & 15 & 17 \\
\texttt{Steel fraction of layer} & 0.68 & 0.48 & 0.74 & 0.55 & 0.60 & 0.84 & 0.78 & 0.82 & 0.82 & 0.82 & 0.60 & 0.80 & 0.74 & 0.90 & 0.86 & 0.93 \\
\hline
\end{tabular}%
}
\end{figure}

The optimization process begins with 15 trials spread across the parameter space to initialize the model. Then, an acquisition function picks the next 5 designs to evaluate. New trials are run in batches of 5 until the surrogate model has low uncertainty across the parameter space. The Pareto front contains the set of designs that cannot improve the performance of one objective without reducing the performance on another objective. Figure~\ref{fig:pareto} visualizes the Pareto front for an experiment where the ratio between the steel and scintillator thickness, and the number of layers are the design parameters. We observe that both MuID objectives prefer to maximize the steel thickness. In contrast, low energy neutrons prefer a steel ratio around to 0.6, and high energy neutrons prefer a larger steel ratio, closer to 0.75. For all four objectives, having more layers reduces the necessary amount of steel to reach a given performance.
\section{Conclusion}
We implement machine learning models into the simulation, reconstruction, and optimization of an hKLM for the EIC. The NF based optical photon parameterization presented provides a notable speed up over the default GEANT4 simulation. We also find that the GNN approach yields excellent results for both MuID and calorimetry, exceeding conventional methods. An hKLM at a future experiment can utilize the presented results to select a detector design to fit the physics needs and physical constraints of the specific experiment.

\acknowledgments
This research was supported by grants from the U.S. Department of Energy, Office of Science, Office of Nuclear Physics under contract DE-SC0024505 and contract DE-SC0024478.


\bibliographystyle{JHEP}
\bibliography{biblio.bib}

@article{Wang:2003bm,
  author = "Wang, J. G.",
  editor = "Fonte, P. and Fraga, M. and Ratti, S. P. and Santonico, R.",
  title = "RPC performance at KLM / Belle",
  doi = "10.1016/S0168-9002(03)01335-4",
  journal = "Nucl. Instrum. Meth. A",
  volume = "508",
  pages = "133--136",
  year = "2003"
}

@misc{frank_markus_2018_1464634,
author       = {Frank, Markus and
Gaede, Frank and
Petric, Marko and
Sailer, Andre},
title        = {AIDASoft/DD4hep},
month        = oct,
year         = 2018,
note         = {webpage: http://dd4hep.cern.ch/},
doi          = {10.5281/zenodo.592244},
url          = {https://doi.org/10.5281/zenodo.592244}
}

@InProceedings{normalizingflows,
  title = 	 {Variational Inference with Normalizing Flows},
  author = 	 {Rezende, Danilo and Mohamed, Shakir},
  booktitle = 	 {Proceedings of the 32nd International Conference on Machine Learning},
  pages = 	 {1530--1538},
  year = 	 {2015},
  editor = 	 {Bach, Francis and Blei, David},
  volume = 	 {37},
  series = 	 {Proceedings of Machine Learning Research},
  address = 	 {Lille, France},
  month = 	 {07--09 Jul},
  publisher =    {PMLR},
  url = 	 {https://proceedings.mlr.press/v37/rezende15.html}
}

@misc{kelleher2026designexpectedperformancehklm,
      title={Design and Expected Performance for an hKLM at the EIC}, 
      author={Rowan Kelleher and Anselm Vossen and William W. Jacobs and Gerard Visser and Simon Schneider and Yordanka Ilieva and Pawel Nadel-Turonski},
      year={2026},
      eprint={2511.08432},
      archivePrefix={arXiv},
      primaryClass={physics.ins-det},
      url={https://arxiv.org/abs/2511.08432}, 
}

@misc{CORE,
      title={CORE -- a COmpact detectoR for the EIC}, 
      author={{CORE Collaboration} and R. Alarcon and M. Baker and V. Baturin and P. Brindza and S. Bueltmann and M. Bukhari and R. Capobianco and E. Christy and S. Diehl and M. Dugger and R. Dupré and R. Dzhygadlo and K. Flood and K. Gnanvo and L. Guo and T. Hayward and M. Hattawy and M. Hoballah and M. Hohlmann and C. E. Hyde and Y. Ilieva and W. W. Jacobs and K. Joo and G. Kalicy and A. Kim and V. Kubarovsky and A. Lehmann and W. Li and D. Marchand and H. Marukyan and M. J. Murray and H. E. Montgomery and V. Morozov and I. Mostafanezhad and A. Movsisyan and E. Munevar and C. Muñoz Camacho and P. Nadel-Turonski and S. Niccolai and K. Peters and A. Prokudin and J. Richards and B. G. Ritchie and U. Shrestha and B. Schmookler and G. Schnell and C. Schwarz and J. Schwiening and P. Schweitzer and P. Simmerling and H. Szumila-Vance and S. Tripathi and N. Trotta and G. Varner and A. Vossen and E. Voutier and N. Wickramaarachchi and N. Zachariou},
      year={2022},
      eprint={2209.00496},
      archivePrefix={arXiv},
      primaryClass={physics.ins-det},
      url={https://arxiv.org/abs/2209.00496}, 
}

@article{diefenthaler2024ai,
  title = "Ai-assisted detector design for the eic (aid2e)",
  author = "Diefenthaler, M. and Fanelli, C. and Gerlach, L. O. and Guan, W. and Horn, T. and Jentsch, A. and Lin, M. and Nagai, K. and Nayak, H. and Pecar, C. and others",
  journal = "JINST",
  volume = "19",
  number = "07",
  pages = "C07001",
  year = "2024"
}

\end{document}